\documentclass[oldversion,a4paper]{aa}
\usepackage{graphicx}
\usepackage{rotating}
\usepackage{txfonts}
\def\teff{$T_{\rm eff}$}
\def\logg{$\log g$}
\def\bz{$\langle B_{\rm z}\rangle$}
\def\vs{$v_{\rm e}\sin i$}
\newcommand{\bs}{$\langle B_{\rm s}\rangle$}
\newcommand{\kms}{km\,s$^{-1}$}

\newcommand{\fifps}[2]{\centering\resizebox{#1}{!}{\includegraphics{#2}}}

\begin{document}

\title{HD 178892 -- a cool Ap star with extremely strong magnetic field}

\author{T.~Ryabchikova\inst{1,2}
\and 
O.~Kochukhov\inst{3}
\and
D.~Kudryavtsev\inst{4}
\and
I.~Romanyuk\inst{4}
\and
E.~Semenko\inst{4}
\and
S.~Bagnulo\inst{5}
\and
G.~Lo Curto\inst{5}
\and
P.~North\inst{6}
\and
M.~Sachkov\inst{2}
}


\institute{Institute for Astronomy, University of Vienna,
T\"urkenschanzstrasse 17, A-1180 Wien, Austria 
\and
Institute of Astronomy, Russian Academy of Sciences, Pyatnitskaya 48, 109017 Moscow, Russia 
\and
Department of Astronomy and Space Physics, Uppsala University Box 515, SE-751 20 Uppsala, Sweden
\and
Special Astrophysical Observatory of RAS, Nizhnij Arkhyz 369167, Russia
\and
European Southern Observatory, Casilla 19001, Santiago 19, Chile
\and
Laboratoire d'Astrophysique, \'Ecole Polytechnique F\'ed\'erale de Lausanne (EPFL), Observatoire, 
1290 Sauverny, Switzerland}

\date{Received / Accepted }

\abstract{We report a discovery of the Zeeman resolved spectral lines, 
corresponding to the extremely large magnetic field modulus \bs\,=\,17.5~kG, in the cool Ap star HD~178892. 
The mean longitudinal field of this star reaches 7.5~kG, and its rotational modulation implies the 
strength of the dipolar magnetic component $B_{\rm p}\ge$\,23~kG.
We have revised rotation period of the star using the All Sky Automated Survey photometry and 
determined $P$\,=\,8.2478~d. Rotation phases of the magnetic
and photometric maxima of the star coincide with each other. We obtained Geneva photometric
observation of HD~178892 and estimated \teff\,=\,7700$\pm$250~K using photometry and the hydrogen Balmer
lines.
Preliminary abundance analysis reveals abundance pattern typical of rapidly oscillating Ap stars.}  

	
\maketitle

\section{Introduction}
\label{intro}

About fifty years have passed since the discovery of the Zeeman resolved line profiles in intensity
spectrum of a magnetic chemically peculiar star (Babcock \cite{B60}). Currently, 47 such objects, with
magnetic field strength in the range of 2.8--33.5~kG, are known (Hubrig et al. \cite{HNS05}). The majority of
the less massive (late A) strongly magnetic stars are slow rotators, with rotation periods of months or
years. However, an exceedingly strong (up to $\sim$8~kG) longitudinal magnetic field, variable with a high
amplitude and a period of about 8.3~days, was recently discovered in the poorly studied SrCrEu Ap star HD~178892 
(BD+$14\degr\,3811$, HIP 94155) by El'kin et al. (\cite{EKR03}) and Kudryavtsev et al. (\cite{KRE04}). 
Such a strong field in combination with a relatively fast
rotation is not known for any other low-mass Ap star, 
thereby making HD~178892 a very interesting
target for in-depth analysis of the atmospheric structure and magnetic field geometry.

The lack of information about HD~178892 motivated us to obtain new magnetic and photometric measurements,
and to acquire the first high-resolution spectroscopic observations for this star.
With these new data,
Zeeman resolved lines indicating a field modulus \bs=17.5~kG are detected in the spectrum of HD~178892,
which confirms the presence of an extremely strong field in this star. We describe new observations of
HD~178892 in Sect.~\ref{observ} and determine fundamental stellar parameters in
Sect.~\ref{parameters}. The stellar 
rotation period and magnetic field are discussed in Sect.~\ref{rotation}. Sect.~\ref{abundances} reports
preliminary abundance analysis results, and conclusions are given in Sect.~\ref{concl}.

\section{Observations and data reduction}
\label{observ}

El'kin et al. (\cite{EKR03}) have selected HD~178892 as a candidate for magnetic measurements on 
the basis of low-resolution spectral observations of the $\lambda$~5200~\AA\ depression profile.
As described by El'kin et al., this low-resolution spectroscopy serves as an analogue of the 
$\Delta a$ photometry and helps to identify strongly magnetic cool Ap stars. 
Circular polarization analyzers (Najdenov \& Chuntonov \cite{NC76}; Chuntonov \cite{C97}) attached to the
Main Stellar spectrograph (MSS) of the 6-m telescope of SAO RAS were used to obtain 18 spectra of HD~178892.
For 16 spectra, the 2K$\times$2K pixel CCD detector was used, and the instrument was configured to give a 
spectral resolution $R$\,=\,15000 and a 240~\AA\
wavelength coverage in the $\lambda$~4500~\AA\ region. Two additional MSS circular
polarization observations
were obtained for the H$\alpha$ region. Polarized spectra were analysed with the help of ESO MIDAS, using our
codes for Zeeman spectra reduction and \bz\ measurements (Kudryavtsev \cite{K00}). The longitudinal component
of the stellar magnetic field, \bz, was determined from the relative shift of the profiles of metal spectral
lines recorded in the left and right circularly polarized spectra. 
Positions of spectral lines were determined using the gaussian approximation of line profiles. Strong magnetic 
field of HD~178892 produces a noticeable partial Zeeman splitting even in low resolution MSS spectra. To exclude the influence of
this effect on the longitudinal field measurements, we selected only lines which were symmetrical
relative to the line centre and did not show Zeeman splitting. This explains a difference of our results with the previous
studies based on the same observations (El'kin et al. \cite{EKR03}; Kudryavtsev et al. \cite{KRE04}), 
where the field was determined using an express procedure without careful selection of spectral lines. Therefore, magnetic field
measurements presented here are more accurate. The log of spectropolarimetric observations and
resulting \bz\ measurements are summarised in Table~\ref{obs}.

High-resolution intensity spectra were obtained for HD~178892 in August 2005 using the cross-dispersed Nasmyth
Echelle Spectrometer (NES, Panchuk et al. \cite{NES}) installed at the SAO 6-m telescope and using the HARPS
spectrograph at the 3.6-m telescope of ESO. The NES spectra, covering the wavelength region
$\lambda\lambda$~4687--6140~\AA\ at the resolving power $R$\,=\,37600--43300, were reduced with the help of the
IDL-based package {\sc reduce} developed by Piskunov \& Valenti (\cite{PV02}).
One HARPS observation in the
$\lambda\lambda$~3780--6910~\AA\ wavelength domain was obtained at
$R$\,=\,115000 and was reduced using the standard pipeline
procedure available for this instrument. Table~\ref{obs} gives information on our high-resolution spectroscopic
observations of HD~178892.

\begin{table}
\caption{Spectroscopic and spectropolarimetric observations of HD~178892.}
\label{obs}
\begin{footnotesize}
\begin{tabular}{crrcc}
\hline
\hline
\multicolumn{5}{c}{Circular polarization spectra}\\ 
HJD &  \bz\ (G)~~ & $S/N$ & $\lambda_{\rm c}$ (\AA) & Instrument  \\
\hline
 2452459.461 &  5940$\pm$520 &  140   & 4500  & MSS \\
 2452625.140 &  7200$\pm$380 &  120   & 4500  & MSS \\
 2452626.139 &  7380$\pm$380 &  100   & 4500  & MSS \\
 2452660.652 &  4400$\pm$710 &   50   & 4500  & MSS \\
 2452661.645 &  3620$\pm$700 &   50   & 4500  & MSS \\
 2452688.614 &  2160$\pm$240 &  130   & 4500  & MSS \\
 2452689.577 &  4290$\pm$480 &  110   & 4500  & MSS \\
 2452805.343 &  5180$\pm$470 &  180   & 4500  & MSS \\
 2452807.392 &  6230$\pm$360 &  140   & 4500  & MSS \\
 2452812.372 &  1460$\pm$290 &  130   & 4500  & MSS \\
 2452830.403 &  5840$\pm$500 &  200   & 4500  & MSS \\
 2452831.435 &  6660$\pm$430 &  200   & 4500  & MSS \\
 2452832.496 &  7170$\pm$390 &  170   & 4500  & MSS \\
 2452834.431 &  4710$\pm$510 &  170   & 4500  & MSS \\
 2452835.380 &  3610$\pm$290 &  240   & 4500  & MSS \\
 2452838.390 &  4900$\pm$500 &  200   & 6560  & MSS \\
 2452840.433 &  6660$\pm$500 &  240   & 6560  & MSS \\
 2453666.163 &  6980$\pm$240 &  240   & 4500  & MSS \\ 

\hline	     		    
\multicolumn{5}{c}{Intensity spectra}\\
HJD & \bs\ (kG) & $S/N$ & $R$ & Instrument\\
\hline
 2453591.362 &17.1$\pm$0.4  &  200 & 43300  & NES\\   
 2453599.357 &17.4$\pm$0.5  &  140 & 37600  & NES\\   
 2453600.263 &17.3$\pm$0.4  &  250 & 37600  & NES\\   
 2453600.575 &18.0$\pm$0.5  &	60 & 115000 & HARPS\\ 
\hline
\end{tabular}
\end{footnotesize}
\end{table}

One photometric measurement of HD~178892 was made in the seven passbands
of the Geneva system (Golay \cite{G80})
with the refurbished photoelectric photometer P7 (Burnet \& Rufener \cite{BR79})
attached to the 1.2-m Mercator telescope at La Palma. The resulting colours
and reddening-free parameters are given in Table~\ref{Geneva}. These data will soon be included
in the up-to-date database maintained by Burki et al. (\cite{B05}), which
complements the Geneva Photometric Catalogue of Rufener (\cite{R88}).

\begin{table*}
\caption{Geneva colour indices of HD 178892, based on a single measurement.
Errors are typically about 0.006 magnitude on each colour. This corresponds to the quality of the
measurement P=2 on the scale defined by Rufener (\cite{R88}).}
\label{Geneva}
\begin{tabular}{ccccccccccc}
\hline
\hline
HJD &V&[U]&[V]&B1&B2&V1&G&B2$-$G&X&Y\\ \hline
2453605.512&8.926&1.396&0.568&0.968&1.391&1.283&1.671&$-$0.280&1.154&$-$0.230\\
\hline
\end{tabular}
\end{table*}

\begin{figure}[t]
\centering\resizebox{7.5cm}{!}{\includegraphics{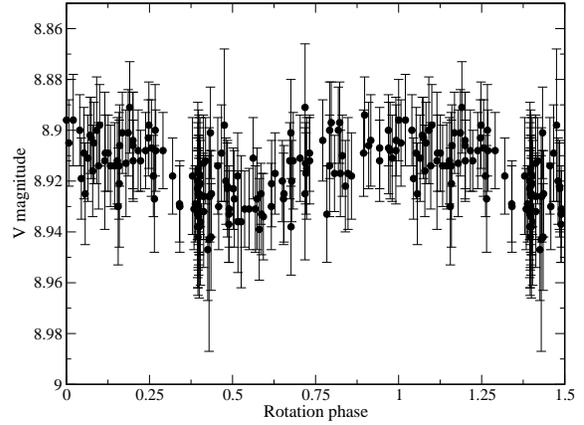}}
\caption{Photometric variation of HD~178892 with a 8.2478~d period.}
\label{asas}
\end{figure}

\section{Fundamental parameters}
\label{parameters}

Since the Geneva reddening-free parameter $Y$ is significantly lower than zero, the star is clearly cooler
than $\sim$\,9000~K (Cramer \& Maeder \cite{CM80}). Thus, one can use the $B2-G$ index to estimate the
effective temperature through the calibration of Hauck \& North (\cite{HN93}). Neglecting interstellar
reddening, this calibration gives $T_{\rm eff}\gtrsim7130$~K. However, the Hipparcos parallax of this star
being 3.70$\pm$1.18~mas, its distance is about $270$~pc, so that reddening may be significant. According to
the maps of Lucke (\cite{L78}) and given the low galactic latitude ($b=+2.78$\degr), the colour excess
$E(B-V)$ may be as large as $\sim0.2$, leading to \teff\,$\sim$\,8440~K. On the
other hand, the 0.11~\AA\
equivalent width of the interstellar Na~D2 line indicates a considerably lower colour excess,
$E(B-V)=0.03$ (Munari \& Zwitter \cite{MZ97}), and results in \teff\,$\sim$\,7300~K. An intermediate
value $E(B-V)=0.1$ would lead to \teff\,$\approx$\,7700~K. 
Although uncertainty 
in the reddening does not allow us to make an accurate temperature estimate from 
the photometry, analysis of the
hydrogen line profiles can decrease substantially the allowed temperature
range. The observed H$\alpha$ and H$\beta$ profiles provide more strict limits for the effective temperature:
7500~K$\le$\teff$\le$ 8000~K. 

The luminosity of HD~178892 is more difficult to determine because of the  low precision of the Hipparcos
parallax. All that can be said, though with much caution, is that HD~178892 appears to be a slightly  evolved
star with $M$\,$\approx$\,1.8--2.4$M_{\sun}$. Assuming the most probable  effective temperature of $7700$~K
and neglecting the Lutz-Kelker correction on the luminosity, interpolation in
the evolutionary tracks of Schaller et al. (\cite{SSM92}) gives $\log
g\approx3.79\pm0.24$. This is smaller than the surface gravity of an unevolved
star (for which $\log g\sim 4.3$), but remains consistent with it within slightly more than $2\sigma$.
With the given uncertainty of the parallax and the effective temperature it is difficult to
make more definite conclusion about the evolutionary status of HD~178892.

\section{Rotation and magnetic field}
\label{rotation}

\begin{figure}[t]
\fifps{7cm}{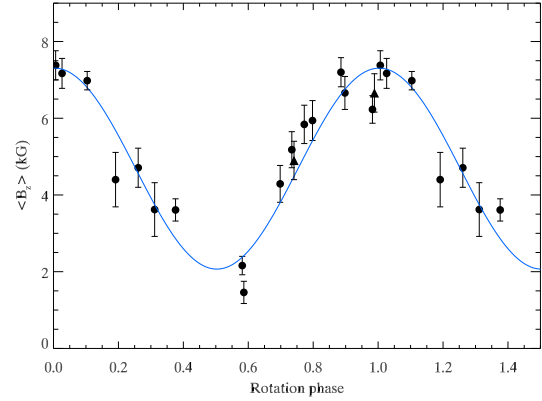}
\caption{Longitudinal field measurements for HD~178892 phased with $P$=8.2478~d. Symbols
show metal line (circles) and H$\alpha$ (triangles) \bz\ determinations.
Solid line shows \bz\ variation expected for the dipolar field topology with $i=36$\degr\ (assumed), 
$B_{\rm p}$\,=\,23~kG, and $\beta$\,=\,37\degr\ (derived from \bz\ observations).}
\label{mag}
\end{figure}

Based on \bz\ measurements Kudryavtsev et al. (\cite{KRE04}) estimated the rotation period of HD~178892  to be
$P$\,=\,8.27$\pm$0.08~d. They did not find any periodic variability of the star in the  Hipparcos epoch
photometry (ESA \cite{ESA}). In addition to the Hipparcos mission, HD~178892 was observed  by the All Sky
Automated Survey (ASAS, Pojmanski \cite{asas02}) and by the Northern Sky Variability Survey  (NSVS, Wo\'zniak
et al. \cite{WVA04}). The latter photometry is made in a wide spectral band with the  effective wavelengths
of the $R$ filter, while the ASAS data (150 points) are obtained in the $V$ system over the time interval of 888 days.
A search for  periodicity using the
Fourier Transform technique realized in the {\sc Period04} software package  (Lenz \cite{L04}) does not
reveal any variation in the NSVS data, whereas a clear signal at $P$=8.25~d  appears in the ASAS photometry.
A refined period search with the help of least-squares IDL routines resulted in the following ephemeris for
HD~178892:
$$
HJD(V_{\rm max})=2452708.562 + 8.2478(76)\cdot E
$$
No other statistically significant peaks were found in the power spectrum as well as in the residuals after
subtracting the main frequency. 
Figure~\ref{asas} shows the ASAS $V$-band variation phased with $P$\,=\,8.2478~d.
The variability of the longitudinal 
magnetic field is illustrated in Fig.~\ref{mag}. From the \bz\ curve we
estimate the minimum possible polar field strength to be $B_{\rm p}$\,$\approx$\,23~kG, assuming a dipolar
field  topology. The corresponding inclination angle is $i$\,=\,37\degr, which coincides with $i$\,=\,36$\pm$15\degr\
derived from the usual oblique rotator relation using $P$\,=\,8.2478~d,
\teff\,=\,7700$\pm$300~K, and the Hipparcos
parallax. The magnetic obliquity $\beta$ is 24--55\degr\ for this permitted range of $i$, whereas \bs\ lies in 
between 15.1 and 21.1~kG depending on the exact values of the angles and rotation phase.

Due to substantial magnetic broadening, it is rather difficult to derive an accurate rotational velocity. 
Spectral lines with negligible splitting, e.g., \ion{Fe}{i} 5434 and 5576 \AA\ are weak and blended. 
The fit to the \ion{Fe}{ii} 4491, 4508 \AA\ and \ion{Cr}{ii} 4634 \AA\ lines with Land\'e factors 
0.4--0.5, along with the \ion{Fe}{i} 5434 \AA\ line allows us to estimate \vs=9$\pm$1 \kms\ as the most probable 
value of the projected rotational velocity. 

\begin{figure}[t]
\centering\resizebox{7cm}{!}{\includegraphics{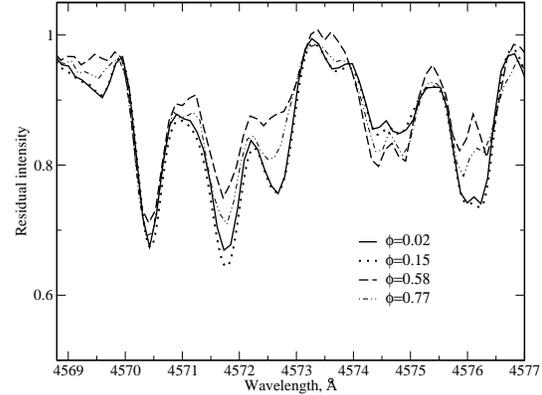}}
\caption{Spectral variations of HD~178892 (MSS observations).}
\label{spvar}
\end{figure}

\begin{figure}[t]
\centering\resizebox{6.8cm}{!}{\rotatebox{-90}{\includegraphics{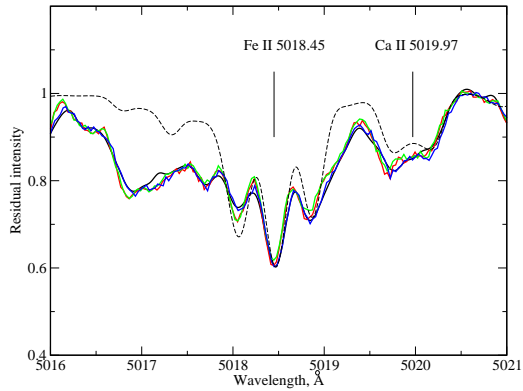}}}
\caption{The spectrum of HD~178892 in the vicinity of the \ion{Fe}{ii} 5018.45~\AA\ line. Solid lines show high-resolution 
observations, whereas dashed line represents synthetic spectrum calculation with \bs=17.5 kG.}
\label{bs}
\end{figure}

HD~178892 is a definite spectral variable (Fig.~\ref{spvar}), but we do not expect to see
significant variations in high-resolution spectra because they all   
are obtained near the maximum of \bz\ in the phase interval 0.040--0.150.
Figure~\ref{bs} displays part of
the spectrum at 5018 \AA, where a triplet-like splitting of the \ion{Fe}{ii} 5018.45~\AA\ line is clearly 
detected in high-resolution spectra. Gaussian approximation for the resolved Zeeman components allow us to measure magnetic splitting 
and deduce the magnetic field modulus \bs, ranging from 17.1$\pm$0.4 kG (1st NES spectrum) to 
18.0$\pm$0.5 kG (HARPS spectrum). The dipolar field geometry derived from the fit to the \bz\ data 
(see Fig.~\ref{mag}) predicts a 18.0--18.5~kG field modulus for the
corresponding rotational phase interval.
Thus, given uncertainty in mutual phasing of the spectropolarimetry and high-resolution intensity 
observations, as well as possible deviation of magnetic topology from a centred dipole, we conclude that
\bz\ and \bs\ measurements agree with each other fairly well.

A synthetic spectrum calculated using {\sc synthmag} code (Piskunov \cite{P99}) with \bs=17.5 kG is compared with 
observations in Fig.~\ref{bs}.

\section{Abundances}
\label{abundances}

Abundances in the atmosphere of HD~178892 were estimated from high-resolution spectra for a model with the parameters \teff=7700 K,
\logg=4.0, and \bs=17.5 kG using the {\sc synthmag} code for magnetic spectrum synthesis. The results 
of this analysis are presented in Table~\ref{abund}. Being derived for the phases near magnetic maximum, they are preliminary
because we did not consider the surface chemical inhomogeneity.  
The atmosphere of HD~178892 is deficient in Mg, Ba,  and
some Fe-peak elements (Ca, Fe, Ni). This is a characteristic signature of cool roAp stars with slightly lower 
temperatures. Abundances of the rare-earth elements (REE) in HD~178892 are also similar to those found in
cool roAp stars. In particular, a strong PrNd anomaly (significant violation  of the ionization equilibrium,
see Ryabchikova et al. \cite{RNW04}) is present. Extremely large  dispersion in abundance obtained from the
resonance and subordinate \ion{Ca}{i} and \ion{Ca}{ii} lines clearly indicate the presence of Ca
stratification. 
Effective temperature variation in the limits provided by the hydrogen line profiles does not result in
substantial changes of the derived chemical abundances. 

\begin{table}
\caption{Element abundances in the atmosphere of HD~178892 derived using $n$ lines. Error estimates are given in
paranthesis. The last column contains new solar abundances from Asplund et al. (2005). }
\label{abund}
\begin{scriptsize}
\begin{tabular}{llrr|llrr}
\hline
\hline
Ion & $\log(N_{\rm el}/N_{\rm tot})$ & $n$ & Sun&Ion & $\log(N_{\rm el}/N_{\rm tot})$ & $n$&Sun\\
\hline
\ion{Mg}{i}   & ~$-$5.20 to $-$5.80  & 3   &-4.51& \ion{Fe}{i}   & ~$-$5.20 to $-$5.50& 5&-4.59 \\
\ion{Mg}{ii}  & ~$-$5.30             & 1   &-4.51& \ion{Fe}{ii}  & ~$-$5.00(10)       & 5&-4.59 \\
\ion{Si}{i}   & ~$-$3.70:            & 1   &-4.53& \ion{Ba}{ii}  & $-$10.30	       & 2&-9.87 \\
\ion{Si}{ii}  & ~$-$3.60(15)         & 4   &-4.53& \ion{Pr}{ii}  & ~$-$8.66(40)       & 3&-11.33\\
\ion{Ca}{i}   & ~$-$6.00 to $-$7.70  & 4   &-5.73& \ion{Pr}{iii} & ~$-$7.44(34)       & 6&-11.33\\
\ion{Ca}{ii}  & ~$-$5.60 to $-$7.70  & 2   &-5.73& \ion{Nd}{ii}  & ~$-$8.05(23)       & 7&-10.59\\
\ion{Sc}{ii}  & ~$-$8.50:            & 1   &-8.99& \ion{Nd}{iii} & ~$-$6.84(33)       &10&-10.59\\
\ion{Cr}{i}   & ~$-$5.50 to $-$6.00  & 6   &-6.40& \ion{Eu}{ii}  & ~$-$8.35(25)       & 4&-11.52\\
\ion{Cr}{ii}  & ~$-$5.00 to $-$5.50  & 6   &-6.40& \ion{Tb}{iii} & ~$-$7.38(32)       & 4&-11.76\\
\hline 
\end{tabular}
\end{scriptsize}
\end{table}
  
\section{Conclusions}
\label{concl}
HD~178892 is a new cool photometric, spectroscopic, and magnetic variable.
Unusually short rotation period of 8.25~d and a very large magnetic field make this star
a perfect target for detailed study of the field geometry and surface chemical inhomogeneities using magnetic Doppler imaging
(Piskunov \& Kochukhov \cite{PK02}). The position of HD~178892 in the H-R diagram
amongst roAp stars, as well as the observed abundance anomalies, make this star a suitable target 
for the search of rapid oscillations. 
New photometric and high-resolution spectropolarimetric observations are needed to improve determination of the rotation period
and to study in details the surface magnetic field and abundance distributions.

\begin{acknowledgements}
We are very grateful to N. Samus for his helpful advises concerning
the automated photometric surveys. We also warmly thank Prof. G. Burki
and an anonymous observer for the photometric measurement in the Geneva
system.
This research was supported by RFBR grant 03-02-16342a, by Russian President's
Foundation for young scientists (MK-1424.2005.2), 
by Austrian Science Fonds (FWF-P17580N2) and by the Swiss
National Science Foundation.
\end{acknowledgements}

\end{document}